\documentclass[11pt]{article}
\usepackage{blois,epsfig}

\def\be{\begin{equation}}
\def\ee{\end{equation}}
\def\bea{\begin{eqnarray}}
\def\eea{\end{eqnarray}}

\begin{document}
\vspace*{4cm}
\title{Top quark physics: From a few to a few millions}

\author{Michele Gallinaro}

\address{
Laborat\'orio de Instrumenta\c{c}\~ao e F\'isica Experimental de Part\'iculas\\
        LIP Lisbon\\ Av. Elias Garcia, 14 - 1000-149 Lisboa - Portugal}

\maketitle\abstracts{
As the heaviest known fundamental particle, the top quark has taken a central role in the study of fundamental interactions. 
The top quark mass is a fundamental 
parameter of the standard model which places constraints on the Higgs boson mass and electroweak symmetry breaking. 
Observations of the relative rates and kinematics of top quark final states may provide constraints for new physics processes. 
Past and current experimental measurements are presented with a critical view, and a look at the future prospects.}

\section{Introduction}

The long search for the top quark engaged researchers	
at laboratories around the world for many years, and came to a successful conclusion 
in February 1995 with the announcement that the top quark had been observed in two experiments at the Tevatron proton-antiproton 
collider at Fermilab~\cite{top_observation}.
The top quark was clearly present in the data and the CDF and D0 collaborations published papers presenting 
overwhelming proof the top quark had been finally found. Its mass was measured to be $m_{t}=176\pm 13$~GeV.
It was expected that the top quark must exist since 1977, when its partner, the bottom quark, was discovered. The weak isospin partner of the bottom quark 
had been missing since then. The top quark completed the three generation structure of the standard model (SM).

Many years after its discovery, 
the top quark still plays a fundamental role in the program of particle physics.
The study of its properties has been extensively carried out in high energy hadron collisions. 
However, a few important questions still remain unanswered. Why is it so heavy? Is its mass generated by the Higgs mechanism? 
What is the role the top quark plays in the electroweak symmetry breaking (EWSB) mechanism?
Does the top quark play a role in non-SM physics? Are the couplings affected?
Since the shutdown of the Tevatron in September 2011, the Large Hadron Collider (LHC) is now the only place 
where it is possible to study the top quark production mechanism.
Before the start of the LHC, most of the measurements were limited by the small number of top quarks available. In comparison, the LHC is a top factory which produced 
one million of top quark events per experiment in 2011 alone. 
The LHC has performed extremely well in the first three years of operation,
and 
the large samples of top quarks collected 
offer the opportunity to improve the results and enter the era of precision measurements and study rare processes. 
It is worth noticing that studies performed at the Tevatron and at the LHC are sometime complementary due to the different energies and production mechanisms.

The top quark is the heaviest of all known elementary particles. 
With a mass close to that of an atom of gold, it has a mass about 40 times heavier than the b-quark and it is heavier than the W boson, 
the top quark decays to 
$t\rightarrow Wb$ with a branching fraction close to 100\%.
The top quark is a fundamental particle with a mass close to the EWSB
scale, and it may play 
an important role in the understanding of the EWSB mechanism.
Furthermore, the large top quark mass implies a large coupling to the Higgs boson, thus establishing a privileged link to the Higgs sector.
Due to its short lifetime, the top quark decays before hadronization and 
it offers the unique opportunity to study the properties of a bare quark which are preserved in the decay chain and transferred to its decay products.
The top quark has charge $Q=+2/3 e$ and weak isospin $T_3=+1/2$, as the partner of the bottom quark in the third generation quark doublet.

At hadron colliders, top quarks are mostly produced in pairs through strong interactions, or individually (single-top production) via electroweak interaction.
Depending on the production mode, the top quark therefore allows different tests of the participating forces.
The top quark is present in the higher-order diagrams and it provides - within the electroweak theory of particle interactions - indirect constraints 
on the Higgs boson mass, together with the W boson mass.
Top quark production also plays an important role in many scenarios in the search for new physics beyond the SM. Several models predict the existence of 
new particles decaying to (or with large couplings to) the top quark. Therefore, the study of top quarks may provide hints to the presence of new physics processes.
Furthermore, top quark production constitutes a large background to many of the searches for new physics processes, and it is therefore important to 
understand the properties and the characteristics of top production and decay mechanisms, and the level of precision will have an impact on the constraints on new physics processes.

For nice and complete reviews on this topic, please see Refs.~\cite{reviews}.

\section{Top quark production}

Several properties of the top quark have already been studied both at the Tevatron and at the LHC. 
These include studies of the kinematical properties of top production, 
measurement of the production cross section, 
reconstruction of $t\bar{t}$ pairs in the fully hadronic final states, study of $\tau$ decays of the top quark, 
and reconstruction of hadronic decays of the W boson from top decays.

\subsection{Top pair production}

At the Tevatron top quark pairs are predominantly produced in quark-antiquark annihilation (90\%), whereas
at the LHC the top quark pair production mechanism is dominated by gluon fusion process ($\simeq 80\%$ at $\sqrt{s}=7$~TeV). 
This is due to the large gluon density in the proton at small-$x$.
The production cross section has been measured in many different final states.
Deviation of the cross section from the predicted SM value may indicate new physics processes.

In each top quark pair event, there are two W bosons and two bottom quarks.
From the experimental point of view, top quark pair events are classified according to the decay mode of the two W bosons:
the all-hadronic final state, in which both W bosons decay into quarks, the lepton+jet final state, in which one W decays leptonically and the other to quarks,
and the dilepton final state, in which both W bosons decay leptonically. 
The word "lepton" here refers to electrons and muons, whereas $\tau$s are somehow classified differently, and they are generally treated separately.
In the dilepton channel, the final state consists of two charged leptons, missing transverse energy, and at least two bottom jets. 
The branching ratio is small (5\%) but the background (mostly Z+jets) is also small, which makes the dilepton the best  final state to select a clean sample of top quark events.
The all-hadronic final state has an experimental signature with at least 6 jets, of which two are from bottom quarks, with a large background, mostly from QCD multi-jet events.
The measurement of the cross section in this channel is rather difficult, despite the large branching fraction (44\%).
The lepton+jet final state offers a compromise with a reasonably large branching fraction (36\%) and a moderate background, mostly from W+jet events. 
The signature consists of one charged lepton, missing transverse energy, and at least four jets (two of them b-jets).

Cross section measurements have been performed both at the Tevatron and at the LHC and the accuracy of the experimental 
results rivals that of theory expectations~\cite{ttbar_xsec_theory}.
At the Tevatron ($\sqrt{s}=1.96$~TeV), the top quark pair production cross section was measured very precisely as
$\sigma=7.50\pm 0.48(\mathrm{stat.+syst.}) $pb~\cite{tevxsec} (6\% precision).
The first top quark pair candidates at the LHC were already reported in the summer of 2010, after a few months of data-taking at $\sqrt{s}=7$~TeV. 
After two years, thousands of top quark events have already been selected.
Measurement of the inclusive top quark pair production cross section have been performed at the LHC in the dilepton and lepton+jet 
channels using electrons and muons and provide the most precise results. 
Most of the cross section measurements are already limited by the systematic uncertainties.
As an example, the cross section in the lepton+jet final state is determined with a simultaneous maximum likelihood fit to the number of jets, the number of b-tagged jets 
and the invariant mass of the tracks associated with the secondary vertex (to allow discriminating light and heavy quark contributions). 
The simultaneous fit in the jet and b-jet multiplicities allow constraining the top quark pair signal and the W+light (and heavy) flavour composition of the background.

Measurement are also performed in the tau+lepton, tau+jets, and all-hadronic channels.
The interest of determining the cross section in all channels is mainly to check the consistency of the measurements, and check for deviations.
For example, the measurement of the cross section in the tau+lepton (as well as the tau+jets) final state is important because a deviation 
of the measured cross section from the expected value may provide a hint for new physics. 
The tau+lepton channel, i.e. $t\bar{t} \rightarrow (\ell\nu_{\ell})(\tau\nu_{\tau})b\bar{b}$ (with $\ell~ =~ e,~\mu$)
is of particular interest because the existence of a charged Higgs with a mass smaller 
than the top quark mass $m_H < m_{t}$ could give rise to anomalous tau lepton production directly observable in this decay channel, via $t\rightarrow H^+ b$. 
As in the other channels, the tau+lepton cross section results~\cite{taudilepton} are consistent with the cross sections measured in the other final states, and
the measurement can be used to set stringent limits on charged Higgs production~\cite{chargedhiggs}.

\subsection{Differential distributions}

The large number of top quark events collected makes also possible the measurement of differential cross sections, $d\sigma/dX$, for the relevant variable $X$.
For instance, variables of relevance may be related to the kinematics of the top or $t\bar{t}$ systems, such as $p_T, M_{t\bar{t}}$, $t\bar{t}+N$~jets. 
These distributions may be used to validate given MC models as well as to check specific higher order QCD calculations. 
Deviations could signal contribution from new physics.
Differential measurements are performed in the dilepton and lepton+jet channels, after reconstruction of the event kinematics.
The measurement of the $t\bar{t}+N$~jet distribution assesses the theoretical predictions and the simulation
in the recoil of the $t\bar{t}$ system and the modelling of additional quark and gluon radiation in $t\bar{t}$ production.
Experimental data are needed to validate the simulated samples and to reduce the uncertainties.
The overall agreement between data and simulation is remarkable, although the uncertainty is not yet at the level to be able to distinguish between different signal models.
The electroweak couplings of the top quark can also be studied in the associated production to a gauge boson, such as $t\bar{t}\gamma, t\bar{t}W$, and $t\bar{t}Z$ events.
Among those, the $t\bar{t}\gamma$ production is large enough to be measured already with the available data samples. The leading-order (LO) cross section for this process 
is about 1~pb at $\sqrt{s}=7$~TeV, for a photon transverse momentum $p_T>8$GeV.
In this process, the photon is radiated from off-shell top quarks or incoming partons, or from on-shell top quark or one of its decay products (such as the W).
The measurement in the lepton+jet channel is performed with a template fit to signal and background, where the templates are derived from data.
Results agree well with SM predictions.

\section{Top quark mass}

The top quark mass $m_t$ is a fundamental parameter of the SM, and it is linked to the W and Higgs boson masses. Through its measurement it 
is possible to constrain indirectly, together with the W mass, the Higgs boson mass value.

It is measured precisely at the Tevatron with an accuracy of about 0.5\%. The combined value from the CDF and D0 experiments yields $m_t=173.2\pm 0.9$~GeV~\cite{mass_tevatron}.
Direct measurements of $m_t$ are also performed at the LHC. Thanks to the large samples of top quarks available, stringent selections and improved analysis techniques, 
and to an excellent performance and good understanding of the detectors, the precision of the LHC measurements are close to the precision reached at the Tevatron 
already after the first few years of data-taking~\cite{mass_lhc}.
The mass is measured in the dilepton, lepton+jet, and in the all-hadronic channels. 
The first measurement at the LHC was performed in the dilepton channel~\cite{mass_lhc_dilepton} from the kinematic characteristics of the events with a full kinematic analysis, 
and with an analytical matrix weighting technique using distributions derived from simulated samples.
The reconstruction of $m_t$ from dilepton events leads to an under-constrained system, since the dilepton channel contains at least two neutrinos in the final state.
The lepton+jet channel provides a fully constrained system, and it is (so far) the "golden" channel as it yields the best accuracy in the mass measurement among all final states.
Many techniques have been used, and the most accurate single measurement at the LHC is performed with the "Ideogram" method, in which a constrained kinematic fit is performed 
for all jet-parton assignment combinations. For each event, a likelihood is calculated as function of the top mass (with two terms, one for signal and one for background) 
corresponding to the probability for the event to be either signal or background. The signal and background probabilities are parametrized using analytic functions, derived from simulation.
An overall likelihood is constructed by multiplying all event likelihoods. 
In general, the mass measurements are limited by systematic uncertainties, and the dominant source is the jet energy scale uncertainty, 
i.e. the absolute scale, ISR/FSR, fragmentation, and single particle response in the calorimeter.
In the lepton+jet channel, reduced uncertainty can be achieved with an {\it in-situ} calibration of the W mass from the untagged jets, using the $W\rightarrow q q'$ decays.

Direct measurements of $m_t$ rely on the reconstruction of kinematic observables sensitive to $m_t$.  
These direct measurements depend on the details of the kinematics, reconstruction, and calibration.
Furthermore, the measurement is performed in a particular definition of $m_t$ which does not correspond to a specific renormalization scheme.
Alternatively, $m_t$ can be derived indirectly from the cross section measurement.
Therefore, measurements of the cross section are used to extract $m_t$ in a well-defined renormalization scheme, such as the pole mass ($m_{pole}$) or $\overline{MS}$ definitions.
The measured inclusive $t\bar{t}$ production cross section is compared with fully inclusive higher-order perturbative QCD computations where 
the top quark mass parameter is unambiguously defined.
For instance, the extraction of $m_{pole}$ from the measured $t\bar{t}$ cross section provides complementary information
compared to direct methods that rely explicitly on the details of the kinematic mass reconstruction.
This extraction also tests the internal consistency of perturbative QCD calculations in a well-defined renormalization scheme, 
and provides an important cross check of the direct measurements.

Direct measurement of a mass difference between particle and anti-particle would indicate a violation of the CPT symmetry.
Quarks carry color charge and cannot be observed directly as they hadronize to colorless particles before decaying. 
One exception is the top quark, as it decays before hadronization due to its short lifetime. 
In the measurement, carried out in the lepton+jet channel, most of the systematic uncertainties cancel out.
The mass difference between top and anti-top quarks $\Delta m_t$ is measured and no significant deviation from zero is found.

\section{Single top production}
 
Single top is produced through electroweak interactions in the s- and t-channels, and in association with a W boson (tW-channel).
With a cross section which is a fraction ($\approx 50\%$) of the top quark pair production, single top has a much larger relative background, and it is therefore more complicated to identify.
Due to the relatively large background, it was first observed at the Tevatron only in 2009 with the help of multi-variate analyses~\cite{tevatron_singletop}.

The dominant production mode is through the t-channel both at the Tevatron and at the LHC; it has the cleanest signature with a light quark jet recoiling against the top quark. 
It is characterized by one central isolated lepton and missing transverse energy, a b-jet and a forward high-$p_T$ recoiling jet. The main background is due to QCD multi-jet and W+jet production.
Single top production in the tW associated production cross section is smaller and it can be studied in the dilepton channel where both W bosons decay 
into a charged lepton and a neutrino, or in the lepton+jet channel. The tW cross section is relatively larger at the LHC than at the Tevatron because of the 
larger gluon fraction in the initial partons.
Among the distinctive features of the tW channel are the presence of a single b-jet,
and the balance between the top and the W.
Even though the observation of the single top production at the Tevatron was based on a combination of t- and s-channels, the s-channel has not yet been observed individually.
With a very small cross section, the s-channel is rather interesting 
as it is sensitive to various new physics processes. 
In particular, it is directly proportional to the Cabibbo-Kobayashi-Maskawa (CKM) matrix element $|V_{tb}|$ 
and it is sensitive to the presence of a $W'$ boson or flavor changing neutral current processes.
At leading order, the final state consists of a top
and a bottom quark.
With the exception of the s-channel, 
which so far has escaped direct detection,
the agreement between experimental results and theoretical predictions is remarkably good.

\section{Top quark properties}

Properties of the top quark are not only interesting to better characterize this fundamental particle, but could also give indication of new physics.
Measurement of the top quark charge, the ratio of branching fractions R=BR($t\rightarrow Wb$)/BR($t\rightarrow Wq$), top quark pair spin correlation,
W polarization in top decays, asymmetry in $t\bar{t}$ 
events are some interesting properties.

\subsection{Measurement of R}

In the SM, the top quark is expected to decay as $t\rightarrow Wb$ with a branching fraction close to 100\%, as top quark decays to a W boson and a quark 
of different isospin are strongly suppressed. 
The magnitude of the CKM matrix element $|V_{tb}|$ is expected to be close to unity as a consequence of unitarity and 
a deviation from this prediction could arise from a fourth quark generation, or simply due to different decay modes. 
Both CDF and D0 experiments have measured R = BR($t\rightarrow Wb$)/BR($t\rightarrow Wq$), where BR($t\rightarrow Wq$) is the branching 
fraction of the top quark to a W boson and a q quark (q=b, s, d). A recent result from the D0 experiment indicates some tension between the SM prediction and
the data, in particular for the dilepton channel where both W bosons decay leptonically.
At the LHC, the measurement is performed by analyzing the b-tagging jet multiplicity in $t\bar{t}$ dilepton events which are expected to be pure in $t\bar{t}$ signal. 
The residual background and the model for the measured b-tag multiplicity are derived from data and account not only for b-tag and mistag probabilities, 
but also include the probability to fully reconstruct the decay products from top quark decays. Results are in good agreement with the SM prediction, 
R=$0.98\pm0.04$(stat.+syst.)~\cite{lhc_vtb}, with a lower limit R$>$0.85 at 95\%CL using the Feldman-Cousins approach.

\subsection{$t\bar{t}$ spin correlation}

In $t\bar{t}$ production, top quarks are unpolarized but their spins are correlated. Due to the short lifetime of the top quark, which is smaller than the hadronization scale, 
the information of the spin correlation is preserved in the decay products. 
It is possible to measure the spin correlation of the top quark pair from the angular correlation of the decay products. Many models of new physics 
predict different 
angular distributions from the SM predictions.
Because of differences in the production mechanisms ($p\overline{p}$ at the Tevatron vs. $pp$ at the LHC) and energy, 
spin correlation measurements are complementary at the Tevatron and at the LHC.
Charged leptons and down-type quarks are most sensitive to the spin correlation measurements. 
However, due to the difficult experimental separation between up- and down-type quarks, spin correlations are measured using dilepton events.
In particular, the correlation coefficient is defined as the fractional difference in the number of events with the spins of the top quarks correlated and those with the spin anti-correlated, 
$A=\frac{N(\uparrow\uparrow)+N(\downarrow\downarrow)-N(\uparrow\downarrow)-N(\downarrow\uparrow)}
{N(\uparrow\uparrow)+N(\downarrow\downarrow)+N(\uparrow\downarrow)+N(\downarrow\uparrow)}$.
A recent Tevatron measurement indicates a 3.1$\sigma$ evidence for spin correlation~\cite{tevatron_spincorr}.
The LHC results, obtained for a given reference frame (either "maximal" or "helicity") are in good agreement with the SM predictions 
and exclude the hypothesis of zero spin correlation~\cite{lhc_spincorr}.

\subsection{W polarization}

The measurement of the polarization of the W boson from top quark decays is interesting as new physics may lead to an anomalous Wtb coupling.
Since the spin information is preserved in top decay products and the bottom quark mass is small compared to the top and W masses, the SM predicts the W boson 
to be mostly longitudinally polarized ($F_0\simeq 69\%$) or left-handed ($F_L\simeq 31\%$) through the V-A coupling. 
These fractions may significantly change in the presence of anomalous couplings, and
may be inferred experimentally from the angular distribution between lepton from the W decay and the b-jet from the same top decay.
The observed distribution has to be corrected for detector effects, such as acceptance and resolution, and theoretical predictions have to account for ISR/FSR, among others.
Measurements show good agreement with the SM predictions, and the results are used to set limits~\cite{polarization}.

\subsection{Charge asymmetry in $t\bar{t}$ production}

In $t\bar{t}$ events, the difference in rapidity (or other) distributions of top and anti-top quarks is usually known as charge asymmetry, which is sensitive to new physics models.
For example, it probes perturbative QCD predictions and provides tests of new physics models where top quark pairs are produced through the exchange of new heavy particles, 
such as axigluons with anomalous axial-vector coupling of gluons to quarks,
Z' bosons, or Kaluza-Klein excitations of gluons.

At the Tevatron, top quarks are emitted in the direction of the incoming quark, anti-top quarks in the direction of the incoming anti-quark. 
Therefore, due to the asymmetric initial state, the asymmetry manifests as a forward-backward asymmetry in the rapidity difference $\Delta y$ between top and anti-top quarks as
$A=\frac{N^+-N^-}{N^++N^-}$, where $N^+ (N^-)$ is the number of events with positive (negative) values of $\Delta y$. 
Recent Tevatron results~\cite{tevatron_asymmetry} yield values larger 
(a $3\sigma$ discrepancy) than the SM predictions.
At the LHC, in $pp$ collisions, there is no forward-backward asymmetry as the initial state is symmetric. The quantity of interest is the charge asymmetry and it shows 
as a preferential production of top quark quarks in the forward direction due to the fact that the anti-quarks  (from the proton's sea) carry a lower momentum fraction. 
Differential measurements have been obtained as a function of $p_T$, $y$, and invariant mass $M_{t\bar{t}}$ of the top quark pair. Measurements are compatible with the SM predictions~\cite{lhc_asymmetry}.

\section{Search for new physics using top quarks}

Top quarks are present in many models of new physics beyond the SM. Some examples include new particles decaying into top quark pairs, 
flavor changing neutral currents, anomalous missing transverse energy, same-sign top pair production, charged Higgs production.

\subsection{Top quark pair invariant mass distribution}

Many extensions to the SM predict interactions with enhanced coupling to the top quarks, resulting as resonances in the $t\bar{t}$ pairs. 
Absence of a resonance in the first two generations is not significant as the coupling could be small in that case.
New particles could be spin-0 scalars or pseudo-scalars, or spin-1 vector or axial-vector particles, such as a Z' boson, a Kaluza-Klein gluon or axigluon, or also spin-2 particles.
Specific analysis tools are developed for the searches in the high-mass regions where the top quarks are highly boosted and the decay products tend to be collimated.
Searches spanning from narrow (with a width of 1-3\%) to wide ($\simeq 10\%$) resonances, both in the low- (up to 1~TeV) and in the high-mass (up to several TeV) regions, 
result in exclusion limits for new particle production rate.

\subsection{Flavor Changing Neutral Currents}

Searches for rare decays of top quarks are possible thanks to the large number of top events collected. 
Top quarks decay to W boson and a bottom quark with a branching fraction of about 100\%. However, some extensions of the SM predict that 
the top quark may also decay to a Z boson and a quark, $t\rightarrow Zq$, where $q$ is a $u$ or a $c$ quark. 
The latter is a decay predicted with a small branching fraction of the order of $10^{-14}$, which is beyond the current experimental reach. 
Therefore, detection of a signal could indicate deviations from the SM predictions.
Search for FCNC processes is sought in the tri-lepton final state, where one top $t\rightarrow Z q\rightarrow \ell\ell q$ is produced.
FCNC can also be sought in the single top production; however, this is experimentally very challenging as the final signature 
$qg\rightarrow t\rightarrow W(\rightarrow \ell\nu) b$ has one isolated lepton, one b-jet, and missing transverse energy, with a large background from W+jet events.
Measurements indicate that branching fractions BR$(t\rightarrow Zq)>0.24\%$ are excluded at 95\%CL.

\subsection{Same-sign top quark production}

New models put forth to explain the larger-than-expected forward-backward asymmetry measurement at the Tevatron require FCNC
in the top sector mediated by the t-channel exchange of a new massive Z' boson.
These mechanisms would generate same-sign top quark pair production. 
However, the LHC results disfavor the region of parameter space consistent with the Tevatron $A_{FB}$ measurement.

\section{Summary and outlook}

After almost twenty years since its discovery, top quark still remains an interesting probe of the SM. 
Large samples of top quarks have allowed detailed tests of the SM up to levels of precision which are challenging the theory. 
However, many fundamental questions still remain, and some of the difficult questions are yet unanswered. 
Some measurements are rapidly approaching (or already did) the grey zone of being limited by systematic uncertainties.

With even larger data samples, additional studies may shed light on important open questions.
Associated production of a Higgs boson with the top quark pair $t\bar{t}H$ is important as it would provide direct determination of the top-Higgs couplings,
separately for the different Higgs decay modes.
Supersymmetry could also affect top quark production. 
Due to the large top quark mass, the lightest scalar top quark $\tilde{t}_1$ can be the lightest scalar quark and even lighter than the top quark itself.
Presence of light top and bottom squarks, charginos and neutralinos could alter the predicted rates through direct stop pair production, 
through the processes $\tilde{t}_1\rightarrow \tilde{\chi}_1^+ b$ or $\tilde{t}_1\rightarrow t \tilde{\chi}^0_1$.
For instance, for a scalar top quark lighter than the top quark, the decay channel $\tilde{t}_1\rightarrow  \tilde{\chi}_1^+(\rightarrow \tilde{\chi}^0_1 W) b$ 
has a similar signature to $t\bar{t}$ events apart from the presence of the neutralinos, whose experimental signature mimics that of a neutrino.

\section*{Acknowledgments}

To my friends and colleagues both at the Tevatron and at the LHC who contributed to producing an incredible wealth of results in the past almost twenty years, 
to those who were present at the time of the discovery and before. 
A special thanks to the organizers of the Blois workshop where I found a wonderful atmosphere in an amazing environment 
which allowed stimulating discussions and exchange of ideas.
To Nature which makes things so hard for us to understand.

\end{document}